\documentstyle[11pt,newpasp,twoside]{article}
\newcommand{\Moy}{\mbox{M$_{\odot}$ yr$^{-1}$}~}

\def\edcomment#1{\iffalse\marginpar{\raggedright\sl#1\/}\else\relax\fi}
\reversemarginpar

\begin{document}
\title{MHD Models of Planetary Nebulae: Review}
 \author{Guillermo Garc\'{\i}a-Segura}
\affil{Instituto de Astronom\'{\i}a-UNAM, Apdo Postal 877,
Ensenada, 22800 Baja California, Mexico}

%
%

\section{Introduction}

When we discuss about MHD effects in planetary nebulae (PNe),
it naturally appears a basic question, which magnetic field do we
study ? One possibility is the ISM magnetic field (e.g. Heiligman 1980),
even more if we are concerned with moving PNe (e.g. Soker \& Dgani 1997).
The next possibility is the internal or stellar magnetic field (Gurzadian 1962).
It is important to start this review by quoting Aller (1958):
{\it ``It has been pointed out by Minkowski and others that the
structural appearance of many planetary nebulae strongly suggest the
presence of magnetic fields. It seems unlikely that such magnetic 
fields are produced} {\em ab initio} {\it in the nebular shell.
Rather, they must have existed in the outer envelope of the 
parent star. Certain red giants stars with magnetic fields may
evolve in such a way that the expansion of the shell is largely
governed by the presence of such a field. Magnetic effects may 
actually be more important than gas pressure differentials and
radiation pressure in controlling the evolution of a planetary 
nebula '' }.
 
The structural appearance of Planetary Nebulae (PNe) display a rich variety 
of shapes, and have been cataloged in a series of morphological classes: 
bipolar, elliptical, point-symmetric, irregular, spherical and quadrupolar 
(Chu, Jacoby \& Arendt 1987; Schwarz, Corradi \& Melnick 1992; 
Manchado et al. 1996ab). 
In contrast, except for a few cases, dust shells
around AGB stars do not show signs of
asphericity (Bujarrabal \& Alcolea 1991; Kahane \& Jura 1994).
Thus, during the transition from the AGB
to the post-AGB phase, one or more physical processes responsible for the
shape of these objects must be initiated.
The origin of aspherical nebulae still remains as one of the fundamental
problems of PNe formation and evolution
(see reviews by Pottasch 1984 and by Iben 1993).

However, Supernova Remnants (SNRs), which  usually present morphologies
extremely similar to PNe (e.g. Gaensler 1998, Fig. 1, 7, 11), have been
commonly related with magnetic shaping. Obviously, this is due to the
easy detection of magnetic fields in SNRs.
The SNR field has evolved largely during past decades on the light of
the MHD. Pionering work by Rees \& Gunn (1974), followed up by 
more sophisticated models (e.g. Kennel \& Coroniti 1984; Begelman \& Li
1992) arrived to the concept of a pulsar wind bubble, in which, a
relativistic pulsar wind containing a toroidal magnetic field
({\em the fast wind}) forms a surrounding bubble where the relativistic
particles and the magnetic flux are accumulated ({\em the hot bubble}).
Both regions are separated by a strong standing MHD shock ({\em the reverse
shock}). Around the bubble, the material previously ejected by the explosion 
({\em the slow wind}) is compressed in a thin shell ({\em the swept-up shell}).
Both regions are separated by a {\em contact discontinuity}.
The magnetic field inside the pulsar wind bubble is expected to be
predominantly toroidal as a result of winding by the pulsar's rotation.
The reader clearly can see the analogy with the two-wind model
proposed by Kwok, Purton \& Fitzgerlad (1978) for PNe. 
In this sense, the matter ejected in the explosion will be identified
as the superwind of PNe.
Are PNe non-relativistic versions of SNRs ?

Recently,  particular emphasis is being given to the explanation
of aspherical planetary nebulae (PNe) due to the effects of the 
stellar magnetic field (e.g. see reviews by Livio 1997; and by L\'opez 1997,
also this volume). 
In this approach, Chevalier \& Luo (1994) have
explored the effects of a rotating star with a magnetized, fast wind on
the formation of aspherical bubbles, following a two-wind model scenario. 
They obtained aspherical steady-state PNe structures. More recently, with 
2-D MHD simulations, R\'o\.zyczka \& Franco (1996) found that the 
time-dependent evolution of a magnetized shocked wind-region has a very complex behavior, and collimated outflows with jet-like features can be created. The 
3-D computations performed by Garc\'{\i}a-Segura (1997) corroborates that the 
magnetized shocked gas indeed creates jets, and the only difference between
the 2-D and 3-D results is that the collimated outflow in 3-D is, as expected, 
likely subject to kink instabilities. Thus, magnetic 
effects could be responsible for the generation of aspherical PNe
as well as collimated flows (jets) observed in some nebulae.

In this paper, we focus on some features that are addressed by 
several works on MHD. 
In \S~2 we review different scenarios for the origin of magnetized winds.
\S~3 discusses the production of axisymmetric flows.
\S~4 describes the confinement of flows and the production of jets and ansae.
\S~5 address the linearly encreasing kinematics of the collimated outflows,
\S~6  give solutions for the special case of point-symmetric nebulae and
\S~7 address solutions for the periodic shells around PNe and proto-PNe.

\section{The magnetized wind origin}

Four major scenarios with three different types of wind driving mechanisms can
be clearly identify:

\begin{description}

\item{ A. Radiation driven wind + B-field : single or wide binary}

This solution incorporates a magnetic field in the classical theory of
Castor, Abbott \& Klein (1975). It works for single stars, however, it
is also appropriate for wide binary systems.
Analytical solutions can be found in Ignace, Cassinelli \& Bjorkman (1998),
for the limit of weak magnetic fields, in which the fields are unimportant
in accelerating the flow. Works under this scenario include
Chevalier and Luo (1994), R\'o\.zyczka \& Franco (1996), Garc\'{\i}a-Segura
(1997), Garc\'{\i}a-Segura  et al. (1999, 2001), Garc\'{\i}a-Segura
\& L\'opez (2000), Gardiner \& Frank (2001), Dwarkadas et al. (2003). 
Solutions including a dipole magnetic field can be found in 
Matt et al. (2000).

\item{B. Magnetic Field driven wind: single or wide binary}

In this scenario, the wind is driven by magnetic pressure, and is the 
B-field the responsible for the mass-loss rate. The main idea of this scenario
is that the field is generated at the stellar interior, which
emerges by magnetic buoyance to the surface. 
This solution also works for wide binary systems.
Works under this scenario include for example 
Pascoli (1992) and Blackman et al. (2001a).

\item{C. Classical binary model: accretion in the secondary}

The theory of MHD launched winds from accretion disk, where toroidal
magnetic fields become finally dominant (e.g. Contopoulos 1995) 
is applied in this scenario. 
Several solutions differs in the accretion mode:
accretion of the primary wind onto the secondary (e.g. Morris 1987, 
Mastrodemos \& Morris 1998); Roche lobe overflow (e.g. Livio, Salzman, 
\& Shaviv 1979; Livio \& Soker 1988). See also review by
Soker in this volume.

\item{D. Alternative binary model: accretion in the primary after common
envelope evolution}

As in scenario C, magnetized winds are espected to be launched from
an accretion disk. In this case, the disk is formed in the primary star.
Recent works on this scenario are Reyes-Ruiz \& L\'opez (1998) and
Blackman et al.(2001b).

\end{description}

In either cases (scenarios A to D) the result is a magnetized wind, in
which the toroidal component is dominant. The degree of self-collimation
and the appearance of a jet, depends on several parameters.
 
\section{Axisymmetric flows caused by a fast, magnetized wind}
 
Axisymmetric flows can be produced by a magnetized wind with or without
the existence of an equatorial density enhancement (EDE) (Begelman \& Li
1992, Chevalier \& Luo 1994, R\'o\.zyczka \& Franco 1996, 
Garc\'{\i}a-Segura 1997, Garc\'{\i}a-Segura et al. 1999, Matt et al. 2000).
Solutions for the formation of EDEs by stellar rotation 
can be found in Bjorkman and Cassinelli (1993), Ignace et al. (1996,1998),
and by dipole magnetic fields in Matt et al.(2000). 

Since the magnetic field at the surface of a post-AGB star is
transported out by its wind, and because of stellar rotation, 
the magnetic field in the wind is dominated by a toroidal component 
(e.g.  Fig. 1 in Ignace et al.1998).    
The toroidal field carried by the fast wind can constrain the motion of the 
flow and an elliptical or bipolar nebula is produced even if the slow wind is 
spherically symmetric.
The resulting toroidal field has a magnetic
tension associated with it (Chevalier \& Luo 1994; Contopoulos 1995). Thus,
the general effect of the  magnetic tension is the elongation of 
the nebula  in the polar direction. 
The mechanism responsible  for the elongation was described by 
R\'o\.zyczka \& Franco (1996), 
and only the basic points are listed here. First, the outer part of the 
magnetized shocked wind region becomes magnetically rather than thermally 
supported, i.e.  the magnetic energy density becomes larger than the thermal 
energy density. The latter decreases due to the expansion of the nebula 
and due to the fact that work must be done against the tension of the 
toroidal field. 
Second, the tension of the toroidal field slows down the expansion
in the direction perpendicular to the symmetry axis, while the expansion
in the direction parallel to the axis proceeds almost unimpeded. Third, a flow
in the shocked wind region toward the symmetry axis 
is initiated, leading to the formation of stagnation regions at the
rotational axis. Note that at the reverse shock, there is a large 
gradient in the magnetic pressure  because the toroidal field at the
rotation axis is nule in the free expanding wind.

\section{Confinement of flows: jets and ansae by pinch effect}

The interaction of the two winds generates a 
wind-shocked region where the tension of the compressed B-field generates an 
important new feature: a pair of flows are induced 
(one at each hemisphere), that move the gas toward the polar regions and
create a pair of collimated outflows at the poles (R\'o\.zyczka \& Franco 1996).
As a consequence and because of the hoop stress of the toroidal field, 
a flow in the shocked wind region toward the symmetry
axis is initiated and maintained, leading to the formation of stagnation 
regions at the axis and to the formation of jets. 
Note that this mechanism works out similarly to that of magnetic confinement
experiments in plasma laboratories ( e.g. Tokamaks, plasma guns).
The pinch effect is introduced for first time by Pascoli (1992) in the context
of PNe.

The gas that arrives at the polar regions of the nebula can form relatively 
dense regions, depending of the radiative cooling conditions 
(Garc\'{\i}a-Segura \& L\'opez 2000),
which may be identified with jets.
Although the hoop stress is always present in this type of flows,
not always the conditions are in favor to form observable jets. Such is 
the case of fast winds with very low mass-loss rates ($10^{-8}--10^{-9}$ 
\Moy). In such conditions,
the fast wind piles up at the poles, but the radiative cooling, proportional 
to $n^2$, is not sufficiently strong to allow the formation of visible jets,
and ansae-like structures are formed instead.
The result of varying the mass loss rate of the fast wind from large 
values to lower values is quite important for jet formation/detection.
(Garc\'{\i}a-Segura \& L\'opez 2000).

Magnetically confined plasmas by toroidal fields can develop kink, 
sausage or neck instabilities (e.g. Jackson 1962), which can easily lead to
the formation of blobs along the collimated plasma flow, giving the 
appearance of episodic jets.  
Another possibility is the MHD Kelvin-Helmholtz instability (Chandrasekar 1961) 
which, given the conditions of the collimated flow, is very likely to appear 
as well, probably combined with kink and neck instabilities.

\section{ Linearly encreasing kinematics of the collimated outflows }

A comparison of the morphologies in observed nebulae with those resulting from
the computed models provides a first test of the model goodness. 
A second verification can be performed with the kinematics of nebulae and, 
specially, with the fast outflows/jets that have been reported for several PNe 
during this last decade. Examples with fast outflows ($\sim 500$ km s$^{-1}$ or higher) are He 3-1475 (Riera et al.1995) and MyCn 18 (Bryce et al.1997). The
observed large velocity values are hardly explained by invoking only 
hydrodynamical effects (Frank et al. 1996), since they are above the critical 
velocity of $\sim 150$ km s$^{-1}$ for which interstellar shocks become 
adiabatic. Magnetohydrodynamical effects, on the other hand, are very 
efficient in forming these fast collimated outflows, as discussed in 
the previous sections.

A comparison of a MHD computation with  the case of MyCn 18 
(Bryce et al. 1997) is shown in Garc\'{\i}a-Segura et al.(1999).
The computed model agrees quite well with figure 3a in Bryce et al. (1997), 
although the model was not specifically intended to reproduce this nebula. 
The most remarkable feature is 
the approximately linear increase of the expansion velocity along the jet, that
matches the actual echelle observations. This accelerative behavior is produced by the relaxation of the magnetic pressure in the shocked region.

\section{ Solutions for point-symmetry}

A particularly intriguing case in PNs morphologies are
those that display point-symmetric structures. 
At first view, the point-symmetric 
morphological class does not look very important. But, a careful inspection
to the statistically, complete sample of the IAC morphological catalog 
(Manchado et al.1996a) reveals 40 objects with some degree of point-symmetric 
features, which represent  19\% of the total list (215) with a well 
defined morphology.
Note that many of these nebulae are not
classified as point-symmetric. In fact, they have been
well classified inside the categories of bipolars and ellipticals, 
such as the case of the Dumbel nebula. This fact points in the direction that
point-symmetry is a common feature related to any morphological
class, instead of a separate group (Manchado et al.2000; see also 
Guerrero, V\'azquez \& L\'opez 1998). 
Such a large fraction of the sample (19\%) suggest that 
the reason which produces point-symmetry should be very common indeed.
 
The most convincing solutions up to now, for the formation of 
point-symmetric nebulae require the existence of a binary system, 
and a magneto-hydrodynamical collimation of the wind,
either for accretion disk winds (Livio \& Pringle 1996; see also
Mastrodemos \& Morris 1998, 
Reyes-Ruiz \& L\'opez 1998 and Blackman et al.2001b)
or for stellar winds (Garc\'{\i}a-Segura 1997, Garc\'{\i}a-Segura \& 
L\'opez 2000).
 
Since the toroidal magnetic field carried out by the wind, either stellar or
coming from a disk (for the last one see Contopoulos 1995), is always 
perpendicular to the rotation
axis of the central star/disk (see Figure 3 in 
Garc\'{\i}a-Segura et al.2000), 
any kind of misalignment from the axis (wobbling instability, precession,
steady tilt respect to the equatorial density enhancement) 
will produce ``naturally'' a point-symmetric nebula.
In either case, it is easy to imagine the topology of the
magnetic field lines in such  scenarios, i.e., multiple
rings centered and aligned along the ``time-dependent'' 
spin axis of the star/disk.
 
Close binaries are a necessary condition in Livio \& Pringle (1996) , while
wide binaries are a sufficient one in Garc\'{\i}a-Segura (1997) and  
Garc\'{\i}a-Segura \& L\'opez (2000).
As described in Garc\'{\i}a-Segura et al. (2002), close binaries, 
either attached or detached, 
will be in favor of forming bipolars with point-symmetric features,
while wide binaries will be in favor of ellipticals with 
point-symmetric features.

\section{Periodic Shells}

The multiple concentric rings, or arcs, that were recently discovered by 
the Hubble Space Telescope around a handful of planetary nebulae (PNe)
is one of the most puzzling and unexpected results delivered by the HST
(see Kwok, Su, \& Hrivnak 1998; Hrivnak, Kwok \& Su 2001; Terzian \& Hajian
2000 and references 
therein). The best documented case of these systems of faint concentric 
rings (hereafter called HST rings) is displayed by NGC 6543 (Balick, 
Wilson \& Hajian 2000). In reality, they are regularly spaced concentric
shells, indicating quasi-periodic events with time intervals, assuming 
typical expansion velocities of AGB winds, in the range of 500 to 1500 
years.  Soker (2000) made a 
critical review of the mechanisms that have been proposed to explain 
them, and he indicates that mass-loss variations associated with 
solar-like magnetic cycles are perhaps the best alternative for their 
origin. 
 
In a more recent paper, Simis, Icke \& Dominik (2001) discuss a new
non-magnetic alternative, and present detailed 1D hydrodynamical
simulations for the acceleration of a dusty AGB wind. 
The stability of this process, however, is 
difficult to explore at the moment in 2-D or 3-D, and it is unclear if 
the compressions 
can truly create shells or if they only lead to inhomogeneities in these
radiatively driven outflows. Thus, at present, one can consider that dusty
flow oscillations (if they really exist) or solar-like
magnetic cycles are two possible candidates to generate the HST rings.
 
In Soker's (2000) view, the magnetic field plays no direct role in the 
evolution of the AGB wind, however temporal variation in the number of 
magnetic spots would be able to modify the mass-loss rate. The number of cool
spots over the AGB surface, which could be preferred sites for dust formation,
is controlled by the magnetic cycle. Thus, given that the mass-loss is 
driven by radiation pressure on dust grains, the same cycle may also 
regulate periodic variations in the mass-loss rate. In this 
interpretation, the magnetic field is a passive player with no dynamical
effects, and the mass-loss rate simply follows the spot cycle activity. 
In addition, to make a logical association with bipolar PNe, he suggests
that a stronger magnetic activity could be expected from dynamo 
amplification in binary systems. 

Garc\'{\i}a-Segura et al. (2001) developed a different point of view and
explored some of the actual
dynamical effects of a solar-like magnetic cycle. The possibility of a
solar-like magnetic dynamo at the AGB phase has been recently discussed by
Blackman et al. (2001a), and they conclude that dynamo amplification is
likely to operate in rotating AGB stars. A logical extension of this result
is that a solar-like activity, including dynamo and cycles, is also expected
in some AGB stars. 
Garc\'{\i}a-Segura et al. (2001) presented 2(1/2)D
magnetohydrodinamic numerical simulations of the
effects of a solar-like magnetic cycle, with periodic polarity inversions,
in the slow wind of an AGB star. The stellar wind was modeled with a
steady mass-loss at constant velocity. This simple version of a
solar-like cycle, without mass-loss variations, was able to reproduce many
properties of the observed concentric rings. The shells were formed by
pressure oscillations, which drive compressions in the magnetized wind.
These pressure oscillations were due to periodic variations in the field
intensity. The periodicity of the shells, then, was simply a half of the
magnetic cycle since each shell was formed when the magnetic pressure
went to zero during the polarity inversion.
Their results showed the importance of MHD effects in the formation 
of the HST rings and successfully reproduced their main features. This 
indicates that modulated mass-loss episodes are not really necessary to 
generate the rings.

\acknowledgments

I thank my colleges and friends J. Franco, J. A. L\'opez, N. Langer,
M. R\'o\.zyczka, A. Manchado, Y.-H. Chu, Y. Terzian, N. Soker,
M. Peimbert and S. Torres-Peimbert for these six, amazing years of 
motivating work on PNe.

\end{document}